\documentclass{article}

\usepackage{arxiv}

\usepackage[utf8]{inputenc} % allow utf-8 input
\usepackage[T1]{fontenc}    % use 8-bit T1 fonts
\usepackage{hyperref}       % hyperlinks
\usepackage{url}            % simple URL typesetting
\usepackage{booktabs}       % professional-quality tables
\usepackage{amsfonts}       % blackboard math symbols
\usepackage{nicefrac}       % compact symbols for 1/2, etc.
\usepackage{microtype}      % microtypography

\usepackage{graphicx}

\newcommand{\figreftext}[1]{Fig.~#1}
\newcommand{\citep}{\cite}

\title{Status in flux: Unequal alliances can create power vacuums}

\author{
  John Bryden\thanks{Corresponding author John Bryden (john.bryden@rhul.ac.uk).} \\
  School of Biological Sciences\\
  Royal Holloway, University of London\\
  Egham, TW20 0EX\\
  United Kingdon \\
   \texttt{john.bryden@rhul.ac.uk} \\
   \And
   Eric Silverman \\
   MRC/CSO Social and Public Health Sciences Unit\\
   University of Glasgow\\
   Glasgow\\
   United Kingdom \\
   \texttt{eric.silverman@glasgow.ac.uk} \\
   \And
   Simon T.~Powers\\
   School of Computing\\
   Edinburgh Napier University\\
   Edinburgh\\
   United Kingdom \\
   \texttt{S.Powers@napier.ac.uk}
}

\begin{document}
\maketitle

\begin{abstract}
Human groups show a variety of leadership structures from no leader, to changing leaders, to a single long-term leader. When a leader is deposed, the presence of a power vacuum can mean they are often quickly replaced. We lack an explanation of how such phenomena can emerge from simple rules of interaction between individuals. Here, we model transitions between different phases of leadership structure. We find a novel class of group dynamical behaviour where there is a single leader who is quickly replaced when they lose status, demonstrating a power vacuum. The model uses a dynamic network of individuals who non-coercively form and break alliances with one-another, with a key parameter modelling inequality in these alliances. We argue the model can explain transitions in leadership structure in the Neolithic Era from relatively equal hunter-gatherer societies, to groups with chieftains which change over time, to groups with an institutionalised leader on a paternal lineage. Our model demonstrates how these transitions can be explained by the impact of technological developments such as food storage and/or weapons, which meant that alliances became more unequal. In general terms, our approach provides a quantitative understanding of how technology and social norms can affect leadership dynamics and structures.
\end{abstract}

\keywords{Leadership dynamics, network analysis, dynamic coevolutionary networks}

\section{Introduction}

Recent changes of the law in China and Turkey mean that their current leaders (President Xi Jinping and President Recep Tayyip Erdo\^{g}an respectively) may hold onto their positions indefinitely. Conversely, in other countries the leader may change relatively often; for instance the UK has had five different Prime Ministers between the years 2007-2019. Among smaller groups some, such as private companies, have a permanent leader while other groups, such as university departments or social societies, have more transient and flexible leadership structures. 

A notable feature of many human groups is that they often do not explicitly coerce their members to join a hierarchy. Instead, soft power and prestige play a strong role \citep{sapolsky_social_2004,van_vugt_evolutionary_2015}, with status being an abstraction of more tangible material resources such as land, food, weapons or other commodities \citep{lin_social_2002}. Status is then voluntarily conferred upon leaders by those they are allied with \citep{hawley_ontogenesis_1999,henrich_evolution_2001,lin_social_2002,pratto_social_2006,castells_network_2011,fiske_virtuous_2014,henrich_secret_2015,van_vugt_evolutionary_2015}, with many of these alliances being asymmetric \citep{hall_asymmetric_1985,henrich_evolution_2001,henrich_secret_2015,van_vugt_evolutionary_2015,holekamp_aggression_2016}. The member with the highest status usually deemed the leader \citep{hawley_ontogenesis_1999,pratto_social_2006,fiske_virtuous_2014,van_vugt_evolutionary_2015}, creating hierarchical societies. What is not clear, however, is what factors determine why some groups have no leader, others have transient leaders and yet others have relatively permanent leaders.

A notable example of a shift between different forms of leadership dynamics is found from evidence from the Neolithic Era. Before this era, human societies consisted of egalitarian hunter-gatherer groups where material resources such as food were shared relatively equally \citep{jaeggi_reciprocity_2013}
and leadership roles were facultative and of a temporary duration. There followed a transition to sedentary groups where high-status individuals had more resources but leaders still changed relatively regularly \citep{bar-yosef_sedentary_2001}. Finally, hereditary leadership became institutionalised, where the role of a chief was passed down a paternal line, which monopolised most of the resources \citep{earle_how_1997}. Many scholars argue that these shifts were due to social and technological developments, which meant that interactions between individuals became increasingly asymmetric. These asymmetries were likely due to control of agricultural surpluses \citep{carneiro_theory_1970,testart_significance_1982,boone_competition_1992,kaplan_hillard_s._evolutionary_2009}, land \citep{carneiro_theory_1970}, ideologies \citep{cauvin_birth_2001}, or military units and weapons \citep{earle_how_1997}.  In light of this evidence, the model we present here investigates how asymmetry in status interaction can generate the different classes of leadership dynamics observed during the Neolithic Era.

Network analysis has proved to be a useful approach for studying the interactions of members of a population \citep{lin_social_2002,latour_reassembling_2007,castells_network_2011}. Quantitative study of hierarchical networks studied is usually static in nature and they are presented as snapshots in time \citep{padgett_robust_1993,albert_statistical_2002}. However, members of a population will often change their properties and who they interact with over time. Recently, a number of studies have concentrated on the dynamical properties of networks \citep{gross_adaptive_2008,bryden_stability_2011}. To look at the factors behind different types of leadership dynamics, we build on this work to model a dynamic coevolutionary network which incorporates the status of individuals as properties of the nodes and links between individuals representing alliances.

\section{Model}

The model consists of a dynamic network of $n$ individuals and is based on the following five assumptions: 1. individuals are equivalent to one another in every way and are unable to coerce one another to form alliances; 2. each individual has a status level which depends on their associations with others, meaning that status is adjusted according to the status of those who they are linked to; 3. individuals share a proportion of their status amongst those they are linked with; 4. status can be asymmetrically passed from one individual to another; 5. individuals can change who they associate with according to the marginal utility of the associations.

Each individual $i$ maintains a status $s_i$ which translates to how much influence they have within the group. Status acts as a multivariate aggregate of an individual's level of money, prestige (titles, jobs, etc), and ownership (land, valuable resources, etc). We assume that, to participate in society, an individual must maintain relationships where they choose a fixed number (constant for all individuals) of associations with other individuals. Consequently, in our model, each individual is given $\lambda$ unidirectional `outgoing' links which are assigned to others in a network. Individuals can have any number of `incoming' links from others in the network. 

The statuses of the individuals are updated according to their links in the status update stage, and individuals are then given an opportunity to rewire their links in the rewiring stage. The model is then run forward in time to observe the distribution of status and changes of that distribution amongst the individuals. We especially looked for `leader' individuals with high status and checked to see whether they were superceded by other leaders. We ran models until patterns of leadership dynamics had stabilised, or for a substantially long time (up to 5 million time steps) to check that new leaders were extremely unlikely to rise to high status.

{\bf Status update stage:} In the model, a proportion of $r$ of each individual's status is shared amongst each of its links (including both incoming and outgoing links). This formalisation of sharing status amongst links is based on Katz's prestige measure \citep{katz_new_1953}. For each link, we calculate that link's status by adding the status contribution from the outgoing individual (for whom the link is outgoing) to the status contribution from the incoming individual. To model unequal alliances we introduce the \emph{inequality parameter} $q$ which reassigns the link's status back to those linked to unequally. A proportion $q$ of a link's status is assigned to the incoming individual, with a proportion $(1-q)$ remaining with the outgoing individual. In this formulation, the total amount of status in the model is constant, and because every status is initialised at $s_i =1.0$, the sum always equals $n$, i.e., $\sum_i s_i=n$.

{\bf Rewiring phase: } In order to maximise status, each individual determines the outgoing link which is worth the least value to itself.  This link is rewired randomly, with probability $w$, to an individual with whom they are not currently linked. The assumption of random rewiring reflects the fact that an individual is unlikely to be able to tell the value of an association with another until they have formed the link.

\section {Results}

In the following, we will present our analysis of the dynamics that result from the interplay between the processes outlined above. Simulations were run of the model choosing parameter values over a complete range. Depending on the parameters, we either observe relatively equal statuses among the population, or a relatively high status level for one or a few individuals. An example of a typical network with a single dominant individual can be seen in Fig. 1. 

\begin {figure}
  \includegraphics[width=0.5\textwidth]{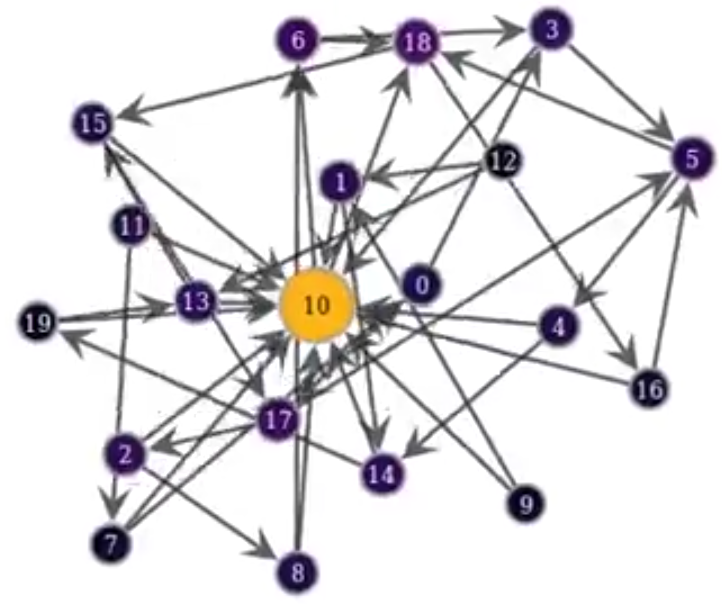}
  \caption{A plot of the network showing a single individual with a high level of status compared to the others. Individuals with higher status are lighter coloured and have larger circles. }
\end {figure}

A key parameter in the model is the inequality parameter ($q$), which models an unequal transfer of status from a link originator to a link receiver. This parameter can represent how social and technological development in the Neolithic Era created inequalities in interactions. As we increase $q$, theory predicts different phases of dynamics in the model, which are shown in \figreftext{2}. We dub the individual with the highest level of status as the leader. We find three different types of leadership dynamics in the model: No leader, transient leader(s), and permanent leader(s).

\begin {figure}
  \includegraphics[width=0.5\textwidth]{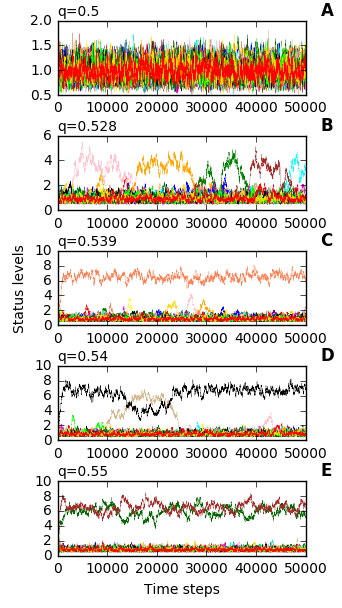}
  \caption{Increasing the inequality takes the model through five different phases. First there is effectively no leader at all (panel A). Then we see that a single individual can rise to a high leadership status, but this is transient and leaders are replaced by other individuals (panel B). The length of time that individuals stay as leader then increases as we increase $q$ until the leader is effectively permanently in charge (panel C). In the next phase, a second individual can rise to a high status alongside the first leader, but these individuals' leadership position is transient (panel D). Finally, two individuals share leadership status and remain so permanently (panel E). 
The value of $q$ is shown, other parameters are $r=0.2$, $n=50$, $\lambda=3$, $w=0.5$. }
\end {figure}

The dynamics shown in Fig. 2 panel B indicate that as one leader is removed, this can create a power vacuum which leads to a new leader quickly gaining status. We have produced a video animation of the model of this phase of the model which is available at www.youtube.com/watch?v=LrQWDhjwsYA. To investigate how the model transitions in and out of this phase, we looked in more detail at how the numbers of leaders and the lengths of time that they were leaders varied over different levels of the inequality parameter ($q$). We calculated the distributions of the different numbers of leaders over a threshold status level ($s_i > 3.0$) and the median length of time new leaders stay over the threshold after they have become the highest status individual. These are plotted in \figreftext{3} with varying numbers of links originating from each individual ($\lambda$). The plot demonstrates there are ranges of the inequality parameter ($q$) where leader turnover is relatively high, but the number of leaders is relatively constant. This shows that there is a power-vacuum effect in our model.

\begin {figure}
  \includegraphics[width=0.5\textwidth]{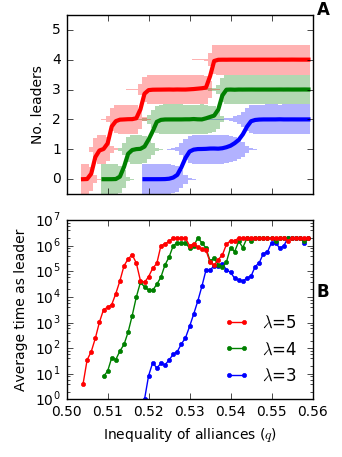}
  \caption{During transient periods, as one leader loses status, another quickly replaces it. Increasing the inequality parameter increases the time that leaders stay above the status level. The lines in panel {\bf A} show the mean numbers of leaders over a threshold status level throughout a complete run of the model. For each value of $q$, vertical boxes indicate the relative proportion of timesteps with each number of leaders, for instance the width of $\approx 0.98$ at $q=0.532$ and $\lambda=3$ indicates that there was generally only one leader above the threshold throughout the simulation. Panel {\bf B} shows the mean time that an individual stays above the threshold after they have reached highest status.Leader turn over was quite high at $q=0.532$ with the average leader lasting 7000 time steps with simulations run over 2 million time steps. Parameters as in \figreftext{2} unless shown.
  }
\end{figure}

To understand the dynamics created by the model in more detail, we compare them to a simpler model of a branching process. Branching processes specify the rates at which an individual with $x_i$ links acquires or loses links. In such processes, we observe a power-law distribution of the number of links when the link gain rate is close to the loss rate. We observe such a power law distribution when this is the case in our model (see Fig. 4, panel A). The formation of a hump in the distribution to the right is also consistent with branching processes with a reflecting boundary. However, the results suggest our model is not simply a superposition of branching processes for each individual. A superposition would not explain why we find a power vacuum effect, where there is always a leader which changes identity over time (see, e.g., \figreftext{3}, $q=0.532$, $\lambda=3$). The reason the power vacuum happens is because the presence of a leader suppresses the others (see \figreftext{4} panel {\bf B}), which is due to the relatively high number of links to the leader.

\begin {figure}
  \includegraphics[width=0.5\textwidth]{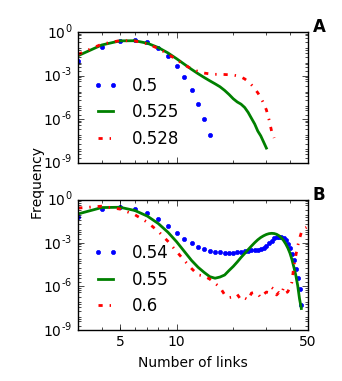}
  \caption{Frequency distributions of the numbers of links to individuals demonstrate how individuals show critical behaviour of a branching process over a range of values of $q$ (panel {\bf A}. There is a truncated power law distribution with exponent $\approx -8.0$ on panel {\bf A} at a relatively low level of $q=0.525$. At higher levels of $q$  we can see how dominant individuals will suppress others to subcritical behaviour (panel {\bf B}). We can see how the rewiring of links to an extra leader between $q=0.54$ and $q=0.55$ (see \figreftext{3} panel {\bf A}) suppresses the frequencies of individuals with mid-range number of links as $q$ is increased. Parameters are the same as in \figreftext{2}, $q$ as shown.  Simulations were run over 2 million time steps.}
\end{figure}

In order to check whether our model is consistent with evidence from the Neolithic Era, we looked to see if the increase in leadership time-length is consistent over a wide range of parameters and that our results can be found in large groups. We ran simulations over a wide range of parameters to investigate the role of each parameter on the dynamics. Over each simulation run, we recorded the number of times there was a change of leader. We counted new leaders when they hadn't been one of the previous $l-1$ leaders. Our results confirm that as we increased the inequality parameter, we recorded fewer leaders as individuals spend longer time periods as leader. The full range of parameters are presented in the Supplementary Material (see Figures S1-9).

\section{Discussion}

The model presented here demonstrates a rich set of leadership structure dynamics amongst individuals in a non-coercive environment. The model reveals an interesting phase where competition to be leader is suppressed by the temporary presence of one leader, meaning that when a leader loses status it will be quickly replaced. We show how the dynamics can depend on the level of inequality of alliances between individuals, and on the numbers of alliances an individual can form. This suggests that technology and social norms can modulate such a system and implies that self-organisation in a society can play a role in keeping a system near to an equilibrium point where leadership changes relatively frequently.

This work pushes forward our understanding of hierarchies in human networks. This is currently largely based on static networks which are formed by preferential-attachment where nodes are more likely to connect to other nodes which are already of high status \citep{albert_statistical_2002}. Our model presents an alternative where the hierarchy is dynamic: nodes have high numbers of connections (alliances) at some points and then other nodes take over. 

This work also contributes new insights into the Neolithic transitions in human societies from relatively flat power structures, through a period where leaders changed over time, to dominant institutionalised leaders \citep{bar-yosef_sedentary_2001}.  Our model presents a potential explanation for this, given that status would be closely linked to control of food or other monopolisable resources. Contemporary to the political transitions were innovations in agriculture, which enabled a high status individual to control a large food surplus. These high status individuals were able to feed a large number of supporters at relatively little cost to themselves, for example funding a military, enabling them to maintain and eventually institutionalise their power.  Monopolisable resources could also be less tangible, such as religious authority \citep{cauvin_birth_2001} however the evidence suggests these changes followed technological advances \citep{whitehouse_complex_2019}. In either case, the growth of population size and the transition to a sedentary, agricultural, lifestyle would have made it more difficult for followers to leave their group and hence easier for a dominant individual to monopolise \citep{carneiro_theory_1970,powers_evolutionary_2014}. These factors, especially the ability to monopolise resources, relate to a high level of the inequality parameter ($q$) in our model as the leader individual is able to form alliances where they exchange a small proportion of these resources to gain loyalty from their supporters.

The three phases of human leadership dynamics correspond to three phases identified in the organisational psychology literature. Lewin has identified three modes of leadership: Laissez Faire, Democratic and Autocratic \citep{lewin_patterns_1939}. In this analysis, the Laissez Faire mode was found to be the case where there is no central resource to coordinate and corresponds to the no-leader phase. When there is a central resource, our model predicts that those individuals with higher status are able to control this central resource, and thus not lose out when more individuals join their group. This is because a controlled surplus of this central resource enables a leader to pay off many individuals and maintain their leadership \citep{bueno_de_mesquita_logic_2005}. The ability to control a central resource means that such groups will switch from the Laissez Faire mode to more Democratic and Autocratic modes. 

In this paper we focussed primarily on applying this model to develop insights regarding the Neolithic transitions from flat power structures to hierarchical societies.  Future work can build upon these foundations to examine whether this model can be applied to other changes in societal structure, such as the movements from monarchy toward parliamentary democracies in 18th-century Europe, or the transitions of Roman civilization betwween monarchy, through annually electing two concurrent consuls in the Roman Republic, to a single Imperator Caesar in the Roman Empire.  Other work might investigate the impact of relaxing some of our assumptions. For instance, exploring different rewiring rules where individuals have different numbers of links, or rewire to others based on a similar or higher levels of status or numbers of links. The model also can be extended in various ways to better represent the real-world contexts in which leadership dynamics operate; these could include representations of technological innovations, changes in social norms, or power struggles between potential leaders.  These extensions would enable us to develop the model further into a powerful exploratory tool for human leadership dynamics. 

\bibliographystyle{unsrt}  
\bibliography{article}  %%% Remove comment to use the external .bib file (using bibtex).

\end{document}